\documentclass[prl,twocolumn,superscriptaddress,showpacs,nobalancelastpage]{revtex4}
\usepackage{graphicx}

\begin{document}

\title{Edge state transport through disordered graphene nanoribbons in the quantum Hall regime}

\author{Fabian Duerr}
\affiliation{Physikalisches Institut (EP3), University of
W\"{u}rzburg, Am Hubland, D-97074, W\"{u}rzburg, Germany}

\author{Jeroen B. Oostinga}
\email{jeroen.oostinga@physik.uni-wuerzburg.de}
\affiliation{Physikalisches Institut (EP3), University of
W\"{u}rzburg, Am Hubland, D-97074, W\"{u}rzburg, Germany}

\author{Charles Gould}
\affiliation{Physikalisches Institut (EP3), University of
W\"{u}rzburg, Am Hubland, D-97074, W\"{u}rzburg, Germany}

\author{Laurens W. Molenkamp}
\affiliation{Physikalisches Institut (EP3), University of
W\"{u}rzburg, Am Hubland, D-97074, W\"{u}rzburg, Germany}

\date{\today}

\begin{abstract}

The presence of strong disorder in graphene nanoribbons yields
low-mobility diffusive transport at high charge densities, whereas a
transport gap occurs at low densities. Here, we investigate the
longitudinal and transverse magnetoresistance of a narrow ($\sim$60
nm) nanoribbon in a six-terminal Hall bar geometry. At $B=$ 11 T,
quantum Hall plateaux appear at $\sigma_{xy}=\pm2e^2/h$, $\pm6e^2/h$
and $\pm10e^2/h$, for which the Landau level spacing is larger than
the Landau level broadening. Interestingly, the transport gap does
not disappear in the quantum Hall regime, when the zero-energy
Landau level is present at the charge neutrality point, implying
that it cannot originate from a lateral confinement gap. At high
charge densities, the longitudinal and Hall resistance exhibit
reproducible fluctuations, which are most pronounced at the
transition regions between Hall plateaux. Bias-dependent
measurements strongly indicate that these fluctuations can be
attributed to phase coherent scattering in the disordered ribbon.

\end{abstract}


\maketitle

Since the discovery of graphene \cite{05-NOV_05-ZHA}, the
investigation of transport through low-dimensional graphene
structures, such as nanoribbons and quantum dots, has obtained much
attention \cite{11-MOL}. Whereas graphene is a gapless material
system, size quantization effects in narrow structures are expected
to open a bandgap \cite{96-NAK_06-BRE_06-SON}. Moreover, depending
on the precise edge structure and interaction effects, peculiar edge
states are predicted to appear \cite{96-NAK_06-BRE_06-SON,
96-FUJ_06_SON_07-RYC_07-TRA}. Transport measurements on ribbons have
shown, however, that the presence of bulk and edge disorder obscures
the observability of bandgap and edge state related transport
properties \cite{11-MOL}.

Strong edge disorder gives rise to enhanced intervalley scattering
in narrow ribbons \cite{EdgeDisorder_Theory}. This results in a
suppression of the carrier mobility in the high-density regime, and
the occurrence of a transport gap in the low-density regime. A
transport gap appears due to the confinement of charges, for which
different origins have been proposed, e.g., the occurrence of a
strong localization regime \cite{10-HAN_10-OOS} or the presence of
disorder-induced electron and hole puddles which are separated by a
bandgap \cite{09-STA_09-TOD_10-GAL}. At high density, the charge
carriers are delocalized and propagate diffusively through a
disordered ribbon. Although transport through magnetoelectric
subbands has been predicted in narrow ribbons with well-defined
edges \cite{06-PER_06-BRE}, experiments have shown that the presence
of strong disorder impedes the occurrence of conductance
quantization at zero magnetic field \cite{08-LIN_10-LIA_11-TOM} and
the development of quantum Hall edge states at high field
\cite{11-RIB_12-MIN_12-HET}. Profound knowledge of disorder effects
on transport is therefore needed to get a better understanding of
the electronic properties of low-dimensional graphene structures.

Here we present a systematic study of electronic transport through a
60 nm wide graphene nanoribbon in a six-terminal Hall bar geometry.
Our transport measurements show that the transport gap at low charge
density does not disappear in the quantum Hall regime, and
therefore, shows that a bandgap due to lateral confinement does not
play a crucial role in the formation of the transport gap at high
magnetic field. Outside the transport gap, the measurements at the
highest applied magnetic field ($B = 11$ T) only reveal the quantum
Hall plateaux at $\sigma_{xy} = \pm 2e^2/h$, $\pm 6e^2/h$ and $\pm
10e^2/h$, for which the corresponding Landau level spacing is larger
than the estimated Landau level broadening. Moreover, the
longitudinal and Hall resistance exhibit aperiodic fluctuations,
which are most pronounced in the transition regions between the
quantum Hall plateaux. Since these reproducible fluctuations are
strongly suppressed when the bias and thermal energy are larger than
the estimated Thouless energy, they can be mainly attributed to
quantum interference effects in the disordered ribbon.

Exfoliated single-layer graphene flakes are transferred to a highly
p-doped Si substrate containing a 285 nm SiO$_2$ top layer. After
deposition of Ti/Au electrodes onto the graphene, the flakes are
etched in an Ar/O$_2$ plasma to obtain narrow nanoribbons (width
$\lesssim 100$ nm) in a six-terminal Hall bar geometry (Fig. 1a).
The etched structures are cleaned at $200^\circ$C in forming gas to
reduce the amount of contamination at the graphene surface. The
transport experiments are carried out in an Oxford dilution
refrigerator at temperatures of 30 mK and 4.2 K. The longitudinal
and transverse resistance are measured as function of current bias
($I_{bias}$), gate-voltage ($V_g$) and magnetic field ($B$) by using
standard lock-in detection techniques. We discuss the data of a
device consisting of a ribbon with length $L = 250$ nm and width $W
= 60$ nm (these data are representative for the other measured
devices of similar dimensions).

\begin{figure}[!t]
\includegraphics[width=3.4in]{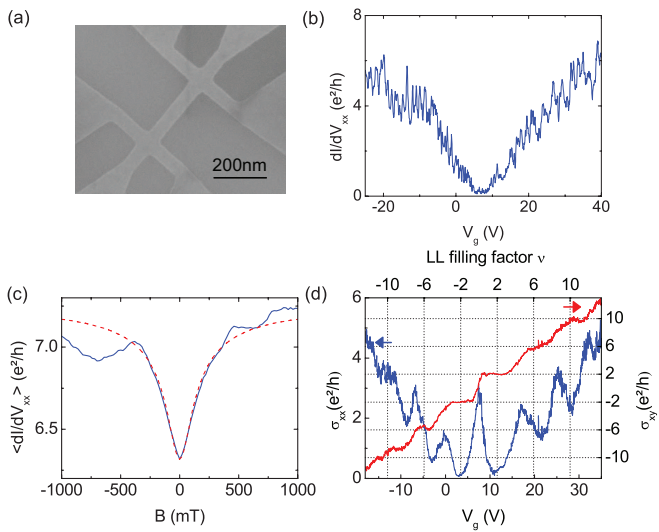}
\caption{(a) Scanning electron micrograph of a six-terminal graphene
nanoribbon device. (b) $dI/dV_{xx}$ versus $V_g$ at $B = 0$ T. (c)
Ensemble average of $dI/dV_{xx}$ versus $B$, determined at high hole
densities in the range of $n \approx$ (4-6)$\cdot 10^{12}$
cm$^{-2}$. The dashed line is a fit of the weak localization theory
(Eq. 12 of Ref. \cite{06-CAN}) to the data with the phase coherence
length as only fitting parameter, yielding $l_\phi \approx 100$ nm.
(d) $\sigma_{xx}$ and $\sigma_{xy}$ versus $V_g$ at $B = 11$ T. The
gate-voltage dependence of the LL filling factor is: $\nu =
\frac{h}{e^2B}C_g (V_g-V_g^D)$. All data are measured at $T = 4.2$
K.} \label{Fig1}
\end{figure}

First we discuss the transport characteristics of the device at $T =
4.2$ K. The differential conductance ($G=dI/dV_{xx}$) versus gate
voltage (Fig. 1b) clearly shows a suppressed conductance at low
density in the vicinity of the Dirac point ($V_g^D \approx 7.5$ V),
as is characteristic for graphene nanoribbon devices \cite{11-MOL}.
At high electron ($V_g > 10$ V) and hole density ($V_g < 5$ V), we
observe reproducible conductance fluctuations. In this regime, the
field-effect mobility of the charge carriers is $\mu =
\frac{L}{WC_g}\frac{dG}{dV_g} \approx 1500$ cm$^2$/Vs where $C_g
\approx 210$ $\mu$F/m$^2$ is the effective gate capacitance
\cite{Note1}.  This corresponds to a diffusion constant of $D =
\frac{L}{W}\frac{G}{\nu e^2} \approx 0.01$ m$^2$/s and a mean free
path of $l_m = 2D/v_f \approx 20$ nm (where $\nu$ is the density of
states and $v_f \approx 10^6$ m/s is the Fermi velocity). Such small
values of the mobility, diffusion constant and mean free path are
typical for plasma-etched graphene nanoribbon devices, and are a
consequence of the high amount of disorder in these systems
\cite{11-MOL, 10-HAN_10-OOS, 09-STA_09-TOD_10-GAL}.

To investigate the occurrence of weak localization in the diffusive
transport regime, we measure the low-field magnetoconductance at
different gate voltages corresponding to high hole densities. Since
each individual measurement at constant charge density exhibits
reproducible conductance fluctuations, we determine the ensemble
average of measurements at different charge densities (Fig. 1c). The
averaged magnetoconductance versus magnetic field clearly shows a
weak localization correction of $\sim$$e^2/h$, indicating the
occurrence of strong intervalley scattering, as expected in
disordered ribbons with short-range disorder at the edges
\cite{06-MOR_08-TIK}. A fit of the weak localization theory for
graphene nanoribbons (i.e., eq. 12 of Ref. \cite{06-CAN}) to our
low-field magnetoconductance data yields a phase coherence length of
$l_\phi \approx 100$ nm (see dashed line in Fig. 1c). This shows
that electronic transport at high carrier densities is in the
diffusive phase coherent regime (i.e., $l_m < L, W \sim l_\phi)$.

Recent experiments on two-terminal graphene nanoribbon devices
\cite{11-RIB_12-MIN_12-HET} have shown that edge state transport
occurs at very high magnetic fields ($B >> 10$ T). In order to study
edge state transport in graphene nanoribbons more profoundly, we
measure the longitudinal ($R_{xx}=dV_{xx}/dI$) and Hall resistance
($R_{xy}=dV_{xy}/dI$) of our six-terminal device at high magnetic
fields up to $B = \pm 11$ T, from which we determine the
longitudinal ($\sigma_{xx}$) and Hall conductivity ($\sigma_{xy}$)
\cite{Note2}. We find that the quantum Hall effect is observable at
magnetic fields larger than $B \approx 6$ T, as expected if $\mu B
\gtrsim 1$. Fig. 1d shows Shubnikov-de Haas (SdH) oscillations of
$\sigma_{xx}$ and quantization steps of $\sigma_{xy}$ as function of
$V_g$ at $B = \pm 11$ T. The Hall plateaux occur at $\sigma_{xy} =
\nu e^2/h$ with $\nu = \pm 2$, $\pm 6$ and $\pm 10$, which
correspond to the characteristic quantization series of Dirac
fermions in single-layer graphene \cite{05-NOV_05-ZHA}.

\begin{figure}[!t]
\includegraphics[width=3.4in]{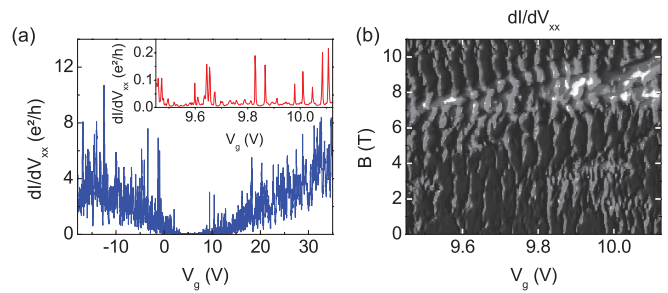}
\caption{(a) $dI/dV_{xx}$ versus $V_g$ at $B = 0$ T. Inset:
$dI/dV_{xx}$ versus $V_g$ in a small $V_g$ range in the transport
gap, showing regions of suppressed $dI/dV_{xx}$ and conductance
peaks. (b) $dI/dV_{xx}$ versus $V_g$ and $B$. The black regions
correspond to suppressed $dI/dV_{xx}$ and the grey/white regions
correspond to regions where $dI/dV_{xx}$ is not suppressed
(conductance peaks). All data are measured at $T = 30$ mK.}
\label{Fig2}
\end{figure}

The measurement data show that the $\nu = \pm 2$ plateaux of
$\sigma_{xy}$ are most pronounced. These plateaux occur when the SdH
oscillations of $\sigma_{xx}$ exhibit a minimum value. However,
$\sigma_{xx}$ does not vanish completely, indicating that scattering
between states at opposite edges is not fully suppressed. Fig. 1d
shows that the $\nu = \pm 6$ and $\nu = \pm 10$ plateaux are less
well developed and do not coincide with the minimum values of the
SdH oscillations of $\sigma_{xx}$. This is probably a consequence of
the strong disorder-induced potential fluctuations in the device
\cite{11-MOL}. The minimum values of $\sigma_{xx}$ increase with
$|\nu|$, indicating an enhancement of scattering between opposite
edges for increasing $|\nu|$. In general, $\sigma_{xy}$ plateaux are
well developed if the energy spacing between the corresponding
Landau levels (LL) is much larger than the LL broadening ($\Gamma$).
In graphene, the LL spacing decreases with increasing $|\nu|$,
whereas the broadening of the LLs does not vary: $\Gamma \approx
\hbar v_f/l_m \approx 30$ meV for each LL. This explains why only
the $|\nu| =$ 2, 6 and 10 plateaux are visible up to $B = 11$ T with
the $\nu = \pm2$ plateaux as the most pronounced ones \cite{Note3}.

Let us continue the analysis by considering the transport
measurements at lower temperature ($T = 30$ mK). In the vicinity of
the Dirac point, the conductance measurement versus $V_g$ at zero
magnetic field clearly shows a transport gap in a gate-range of
$\Delta V_g \approx 8$ V (Fig. 2a). This corresponds to an energy
scale of $\Delta E = \hbar v_f \sqrt{2\pi C_g \Delta V_g/e} \approx
160$ meV, which can be attributed to the disorder-induced potential
fluctuations in the ribbon \cite{10-HAN_10-OOS}, possibly in
conjunction with a lateral confinement gap
\cite{09-STA_09-TOD_10-GAL}. Earlier transport experiments have
shown that, in this low-density regime, electrons are confined in
small areas of the ribbon \cite{11-MOL}. In our conductance
measurements, we clearly observe conductance peaks in the transport
gap (inset of Fig. 2a), showing that electron states are indeed
localized in the low-density regime and transport is dominated by
charging effects. By measuring the conductance as function of
magnetic field in a small gate-voltage range of the transport gap
(Fig. 2b), we observe conductance peaks in the full magnetic field
range, including the quantum Hall regime ($B \gtrsim 6$ T). This
result shows that electrons are confined at low densities, even when
the zero-energy LL appears at the Dirac point. This rules out that a
bandgap due to lateral confinement plays a crucial role in the
physical mechanism underlying the occurrence of the transport gap at
high magnetic field (consistent with \cite{EdgeDisorder_Theory,
10-HAN_10-OOS}), because the zero-energy LL closes such a size
quantization gap \cite{Note4}.

\begin{figure}[!t]
\includegraphics[width=3.4in]{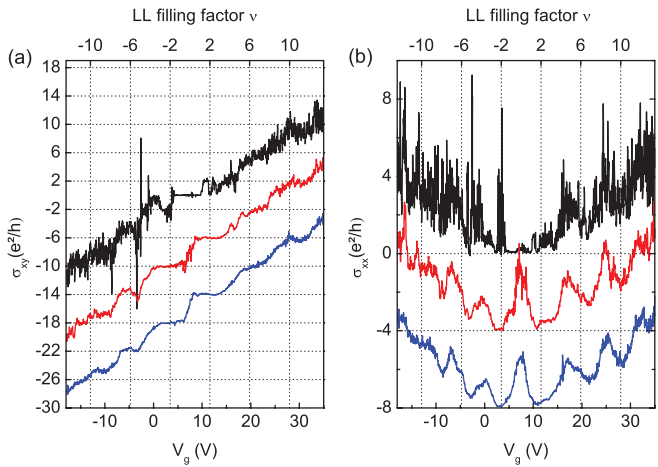}
\caption{(a) $\sigma_{xy}$ and (b) $\sigma_{xx}$ versus $V_g$ at
$B=11$ T and $I_{bias} = 0$ nA (black), 86 nA (red) and 344 nA
(blue). The red and blue curves are shifted with respect to the
black curve for clarity. All data are measured at $T = 30$ mK.}
\label{Fig3}
\end{figure}

\begin{figure}[!t]
\includegraphics[width=3.4in]{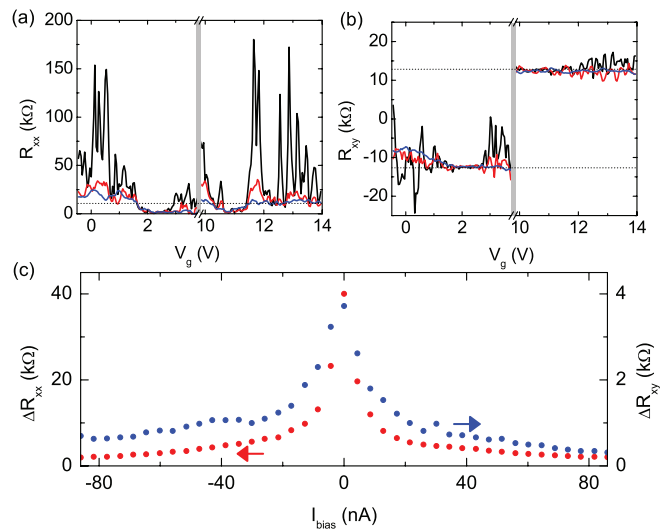}
\caption{(a) $R_{xx}$ and (b) $R_{xy}$ versus $V_g$ at $B= 11$ T and
$I_{bias} = 0$ nA (black), 17 nA (red) and 86 nA (blue). The $V_g$
ranges correspond to the regions of the $\nu=\pm2$ plateaux; see
dotted line in (b). (c) The average fluctuations $\Delta R_{xx}$ and
$\Delta R_{xy}$ versus $I_{bias}$, determined in the $V_g$ range
shown in (a) and (b). All data are measured at $T = 30$ mK.}
\label{Fig4}
\end{figure}

At $T = 30$ mK, we observe large reproducible fluctuations of
$\sigma_{xx}$ and $\sigma_{xy}$ at high charge densities in the
quantum Hall regime (Fig. 3a,b), which obscure the visibility of SdH
oscillations of $\sigma_{xx}$ and quantization plateaux of
$\sigma_{xy}$. The transport gap gives rise to a strongly suppressed
conductivity tensor in the low-density regime ($\sigma_{xx},
\sigma_{xy} \rightarrow  0$ if $R_{xx} \rightarrow \infty$; see
\cite{Note2}). Fig. 3a,b show that the transport gap disappears when
we apply a sufficiently high bias, leading to non-vanishing values
of $\sigma_{xx}$ and $\sigma_{xy}$ in the vicinity of the Dirac
point. Moreover, in the high density regime, the fluctuations of
$\sigma_{xx}$ and $\sigma_{xy}$ are suppressed at high bias, and the
SdH oscillations of $\sigma_{xx}$ and the quantization plateaux of
$\sigma_{xy}$ are more visible.

In order to investigate the origin of the reproducible fluctuations
at low temperature, we analyze the fluctuations of $R_{xx}$ and
$R_{xy}$ in the region of the $\nu = \pm 2$ Hall plateaux (since
these are the best developed plateaux in our measurements). Fig.
4a,b show that the fluctuations ($\Delta R_{xx}$, $\Delta R_{xy}$)
are smallest in the center region of the plateaux, whereas they
increase strongly farther from the center region (i.e. in the
transition regions between adjacent plateaux). The bias-dependence
of $\Delta R_{xx}$ and $\Delta R_{xy}$ (Fig. 4a-c) shows that the
fluctuations in the transition regions are strongly suppressed at
$I_{bias} \gtrsim 10$ nA (this corresponds to $V_{bias} \gtrsim 100$
$\mu$V, because $R_{xx} \sim 10$ k$\Omega$ in the transition
regions; see dotted line in Fig. 4a).

Resistance fluctuations in the quantum Hall regime have been studied
extensively in mesoscopic devices based on Si and GaAs
heterostructures \cite{QHEfluctuations_exp}. When the device
dimensions are of the order of $l_\phi$, phase coherent scattering
mechanisms give rise to resistance fluctuations in the quantum Hall
regime, similar to universal conductance fluctuations in the
zero-field regime \cite{91-BEE_92-BUE}. Since our nanoribbon
dimensions are of the order of $l_\phi$, the observed resistance
fluctuations may originate from quantum interference. Fig. 4c shows
indeed that the fluctuations are suppressed on an energy scale of
$\sim$100 $\mu$eV, which is of the same order as the Thouless energy
of the ribbon ($E_{Th} = \hbar D/L^2 \approx 100$ $\mu$eV). This
also explains why these fluctuations are suppressed at 4.2 K, when
$kT > E_{Th}$. Thus, our results indicate that the observed
fluctuations in the quantum Hall regime can be mainly attributed to
phase coherent scattering mechanisms in the disordered graphene
nanoribbon.

In conclusion, we have measured the quantum Hall effect in a 60 nm
wide graphene nanoribbon, which results in the observation of SdH
oscillations of $\sigma_{xx}$ and quantized plateaux of
$\sigma_{xy}$ for $\nu = \pm 2$, $\pm 6$ and $\pm 10$. At $T = 30$
mK, we observe large fluctuations of $\sigma_{xx}$ and $\sigma_{xy}$
at high charge densities, which may be attributed to phase coherent
scattering mechanisms in the ribbon. At low charge densities, the
electrons are confined in the ribbon yielding a transport gap, in
which transport is dominated by charging effects. Since the
transport gap does not disappear in the quantum Hall regime, a
bandgap due to lateral confinement cannot play a crucial role in the
occurrence of the transport gap at high magnetic field (which
confirms what has been reported before \cite{EdgeDisorder_Theory,
10-HAN_10-OOS}).

\begin{acknowledgments}

We thank H. Hettmansperger, J. Schelter and B. Trauzettel for useful
discussions. This work was financially supported by the German
research foundation DFG [DFG-JST joint research program 'Topological
Electronics'], and the EU ECR Eurocores programme [Eurographene
project].

\end{acknowledgments}

\end{document}